\documentclass[%
 reprint,
superscriptaddress,
 amsmath,amssymb,
 aip,
 sd,
]{revtex4-2}

\usepackage{graphicx}
\usepackage{dcolumn}
\usepackage{bm}

\begin{document}

\title{A kiloelectron-volt ultrafast electron micro-diffraction apparatus \\ using low emittance semiconductor photocathodes}

\author{W. H. Li}
\thanks{These two authors contributed equally.}
\affiliation{%
Cornell Laboratory for Accelerator-Based Sciences and Education, Cornell University, Ithaca, NY 14853, USA}%
\author{C. J. R. Duncan}
\thanks{These two authors contributed equally.}
\affiliation{%
Cornell Laboratory for Accelerator-Based Sciences and Education, Cornell University, Ithaca, NY 14853, USA}%
\author{M. B. Andorf}%
\affiliation{%
Cornell Laboratory for Accelerator-Based Sciences and Education, Cornell University, Ithaca, NY 14853, USA}%
\author{A. C. Bartnik}
\affiliation{%
Cornell Laboratory for Accelerator-Based Sciences and Education, Cornell University, Ithaca, NY 14853, USA}%
\author{E. Bianco}
\affiliation{%
Kavli Institute at Cornell for Nanoscale Science, Ithaca, NY 14853, USA}%
\author{L. Cultrera}
\affiliation{%
Cornell Laboratory for Accelerator-Based Sciences and Education, Cornell University, Ithaca, NY 14853, USA}%
\author{A. Galdi}
\affiliation{%
Cornell Laboratory for Accelerator-Based Sciences and Education, Cornell University, Ithaca, NY 14853, USA}%
\author{M. Gordon}
\affiliation{University of Chicago, Chicago, IL, 60637, USA}
\author{M. Kaemingk}
\affiliation{%
Cornell Laboratory for Accelerator-Based Sciences and Education, Cornell University, Ithaca, NY 14853, USA}%
\author{C. A. Pennington}
\affiliation{%
Cornell Laboratory for Accelerator-Based Sciences and Education, Cornell University, Ithaca, NY 14853, USA}%
\author{L. F. Kourkoutis}
\affiliation{%
Kavli Institute at Cornell for Nanoscale Science, Ithaca, NY 14853, USA}%
\affiliation{%
School of Applied and Engineering Physics, Cornell University, Ithaca, NY 14853, USA}%
\author{I. V. Bazarov}
\affiliation{%
Cornell Laboratory for Accelerator-Based Sciences and Education, Cornell University, Ithaca, NY 14853, USA}%
\author{J. M. Maxson}
\thanks{The authors to whom correspondence may be addressed: jmm586@cornell.edu, whl64@cornell.edu.}
\affiliation{%
Cornell Laboratory for Accelerator-Based Sciences and Education, Cornell University, Ithaca, NY 14853, USA}%
 

\date{\today}

\begin{abstract}
We report the design and performance of a time-resolved electron diffraction apparatus capable of producing intense bunches with simultaneously single digit micron probe size, long coherence length, and $200$ fs rms time resolution. We measure the 5d (peak) beam brightness at the sample location in micro-diffraction mode to be $7 \times 10^{13} \ \mathrm{A}/\text{m}^2\text{-rad}^2$. To generate high brightness electron bunches, the system employs high efficiency, low emittance semiconductor photocathodes driven with a wavelength near the photoemission threshold at a repetition rate up to 250 kHz. We characterize spatial, temporal, and reciprocal space resolution of the apparatus. We perform proof-of-principle measurements of ultrafast heating in single crystal Au samples and compare experimental results with simulations that account for the effects of  multiple-scattering.
\end{abstract}

\maketitle


\section{Introduction}
Probing the transient response of materials after excitation by short, intense pulses of light is a route to the discovery of new phenomena and functionalities not observable in equilibrium \cite{ihee2001direct, cavalleri2001femtosecond, siwick2003atomic, gedik2007nonequilibrium, stojchevska2014ultrafast, sie_ultrafast_2019}.  Progress in the experimental investigation of out-of-equilibrium states of matter requires the ongoing development of electron and x-ray beam-based tools with high resolving power in space, time, and energy. Observation of the fastest structural responses demands temporal resolution at the picosecond level and below \cite{morimoto2018diffraction, maxson2017direct}. 

In practical terms, electron beam systems can be more compact than synchrotron x-ray sources, and can provide complementary structural information \cite{li2017application}. Multiple ultrafast electron probe modalities are common, ranging from diffraction \cite{mo2018heterogeneous} to microscopy (imaging) \cite{cremons2016femtosecond} and spectroscopy \cite{EELS}, accelerated to energies from sub-keV \cite{vogelgesang2018phase} to keV \cite{chatelain2012ultrafast}, up to few MeV \cite{wang2003femto, zhu2015femtosecond, weathersby_mega-electron-volt_2015, yang2009100, ji2019ultrafast, fu2014high}, with notable advantages in each regime. Among these, keV diffraction beamlines have served as the pioneering ultrafast electron structural probe system \cite{mourou1982picosecond, williamson1997clocking, siwick2003atomic, Ischenko1983, Dudek2001}, and keV primary energies offer the advantages of room-scale size, high scattering cross section (beneficial for the investigation of 2D materials) \cite{badali2016ultrafast}, and intrinsically high reciprocal space and energy resolution \cite{vanacore2018attosecond, vanacore2019ultrafast, zong_mirror_2018, pomarico2017mev}. In this paper, we focus on the keV ultrafast electron diffraction (UED) system archetype and explore ways to expand its scientific reach.   

Temporal resolution has rightly been a focus of much work on ultrafast electron probes,  with the state of the art in temporal resolution well below 50 fs \cite{zhao2018terahertz, snively2020femtosecond, qi2020breaking}. In this work, we adopt the rf compression scheme employed by multiple keV sources which regularly achieve 100 fs temporal resolution and below.

In the design of this apparatus, we emphasize the transverse degrees of freedom: probe size and reciprocal space resolution \cite{shen2018femtosecond, uenc}.  Decreasing the probe size can significantly reduce the difficulty and time requirements of sample preparation \cite{bie2021versatile}, may permit the direct usage of standard transmission electron microscope (TEM) sample preparation techniques \cite{langford2008situ}, or can enable selected area ultrafast diffraction of textured materials \cite{zong_mirror_2018}. Additionally, for a given pump fluence, reducing the probe size permits a commensurate reduction of pump size and pump-pulse energy. Reducing the total deposited energy in the sample per shot can both extend sample lifetimes and potentially shorten the time required for the sample to relax to its initial state following pump interaction, allowing data to be taken at higher repetition rates. 

As a means for beam generation, photoemission affords fine control of the electron distribution both in space and time via laser shaping \cite{maxson2014efficient, maxson_adaptive2015, thompson2016suppression, franssen2019compact}. To increase spatial beam quality, in this work we employ photocathodes with high intrinsic brightness and further tune the photoemission wavelength \cite{kasmi2015femtosecond, karkare2020ultracold}. The choice of photo-excitation energies, when approaching the photoemission threshold, involves a trade-off between (i) maximizing the ratio of emitted electrons to incident photons --- the {\em quantum efficiency} (QE), and; (ii) minimizing the momentum spread of emitted photoelectrons, summarized in the {\em mean transverse energy} (MTE) \cite{musumeci_advances_2018} of the beam. The maximum brightness achievable from a photoemission source is inversely proportional to MTE \cite{bazarov2009, daniele_max_current_2014}.

Almost all ultrafast electron machines rely on metal photocathodes. While being robust, metals typically have much lower quantum efficiency (often $<10^{-4}$, but there are exceptions \cite{qian2010surface}) and higher MTE (often 100s of meV) than the best performing semiconductor photocathodes. We elect to use alkali antimonide photocathodes, which possess higher QE ($\sim 10^{-3}$) and lower MTE ($<50$ meV) when illuminated with near-threshold visible and near-infrared photons. This high quantum efficiency near the photoemission threshold is critical as it mitigates the brightness diluting effects of multiphoton photoemission and ultrafast cathode heating \cite{maxson_ultrafast_2017, Bae2018}. However, the cost of this increased brightness is extreme vacuum sensitivity. Unlike metal photocathodes, alkali antimonides must be grown, transported, and used in ultra high vacuum (UHV) conditions.

Multiple ultrafast microscopy sources have utilized sharp tip geometries to generate coherent, sub-micron probes \cite{zhang_observation_2019, danz_ultrafast_2021, houdellier2018development}. Similar sub-micron emission sizes in high quantum efficiency semiconductor cathodes are yet to be demonstrated but are possible in principle. Low MTE photoemission is, in a sense, the conjugate technique to spatial confinement, but has the advantage that it does not restrict total emitted charge per pulse.  In this work, we use a 10 micron diameter laser-machined aperture just upstream of the sample plane (henceforth referred to as the ``probe-defining aperture") to generate few-micron rms sample probe sizes. In the hypothetical case without space charge, the role of the photocathode MTE is straightforward; our electron optics roughly image the photocathode plane onto this aperture, and the MTE then primarily determines the reciprocal space resolution of the diffraction pattern. However, we operate in a regime where space-charge is non-negligible, and this modifies the direct correspondence between MTE and reciprocal space resolution. Space charge forces usually degrade the brightness of the total beam, a fact which on the surface suggests that low MTE may not be useful for beam brightness at the sample. However, for both high-charge photoinjectors (for example, for synchrotron light sources) and UED machines of the kind described in this work, it has been shown in simulation that if beam optics and setpoints are designed from the outset to incorporate low MTE photoemission, significant performance gains are possible, down to the levels of the lowest MTEs measured from photoemission to date: $\sim 10$ meV \cite{pierce2020}. We present below simulations of this effect at play in our apparatus, as well as measurements consistent with the simulations. 

The effects of space charge are further mitigated by selecting only the spatial core of the beam, as we do via the probe-defining aperture. In the core of the beam, nonlinear space charge forces are much smaller than at the edges, so that the core of the beam remains brighter than the beam as a whole \cite{zerbe2018, bazarov2009, pierce2020}. We typically overfill the probe-defining aperture by orders of magnitude in charge so that we select only this core, whose brightness is most closely determined by the initial MTE and photoemitted charge density. As compared to space charge free operation, this method provides significant flux gains through the aperture with only modest concessions in coherence. For probe diameters of 10 micron, we typically select between 100-1000 electrons from incident pulses of $\sim 10^5$ electrons. The resulting bunch length at the sample is $< 200 $ fs rms.

Operated in micro-diffraction mode at its maximum repetition rate of 250 kHz with between 100 to 1000 electrons per pulse, we measure a peak brightness of $7\times 10^{13} \mathrm{A/m^2\text{-rad}^2}$ at the sample plane, which is the state of the art in this charge regime.  Thanks to its photoinjector heritage, our design mitigates much of the performance-degrading effects of space charge and in high-charge mode, at $10^5$ electrons per bunch, achieves a peak brightness of $4\times 10^{13} \mathrm{A/m^2\text{-rad}^2}$, a small decrease in brightness for a dramatic increase in charge. Wide flexibility in bunch charges is advantageous in UED experiments particularly at moderate pump fluences, even in stroboscopic operation, because sample relaxation times and damage thresholds can prohibit running at megahertz repetition rates \cite{chase2016ultrafast, curtis2021toward}. Moreover, the intrinsically lower energy spread of our photocathode makes our machine suitable for performing sub-eV ultrafast electron energy loss spectroscopy experiments in the future. Finally, we compare below our brightness figures with brightness measurements in both MeV and keV ultrafast electron diffraction and microscopy systems, and show that our machine operates in an unfilled, complementary parameter space.

The body of this paper begins in Section \ref{sec:system} by describing the subsystems of the beamline, which we have named MEDUSA (Micro-Electron Diffraction for Ultrafast Structural Analysis). In sections \ref{sec:spatial} and \ref{sec:temporal} we report diagnostic measurements of the spatial and temporal resolution of the electron probe. Section \ref{sec:ued} concludes the paper by presenting the results of a proof-of-principle measurement of ultrafast heating in a gold sample.

\begin{figure*}
    \centering
    \includegraphics[width=1.0\linewidth]{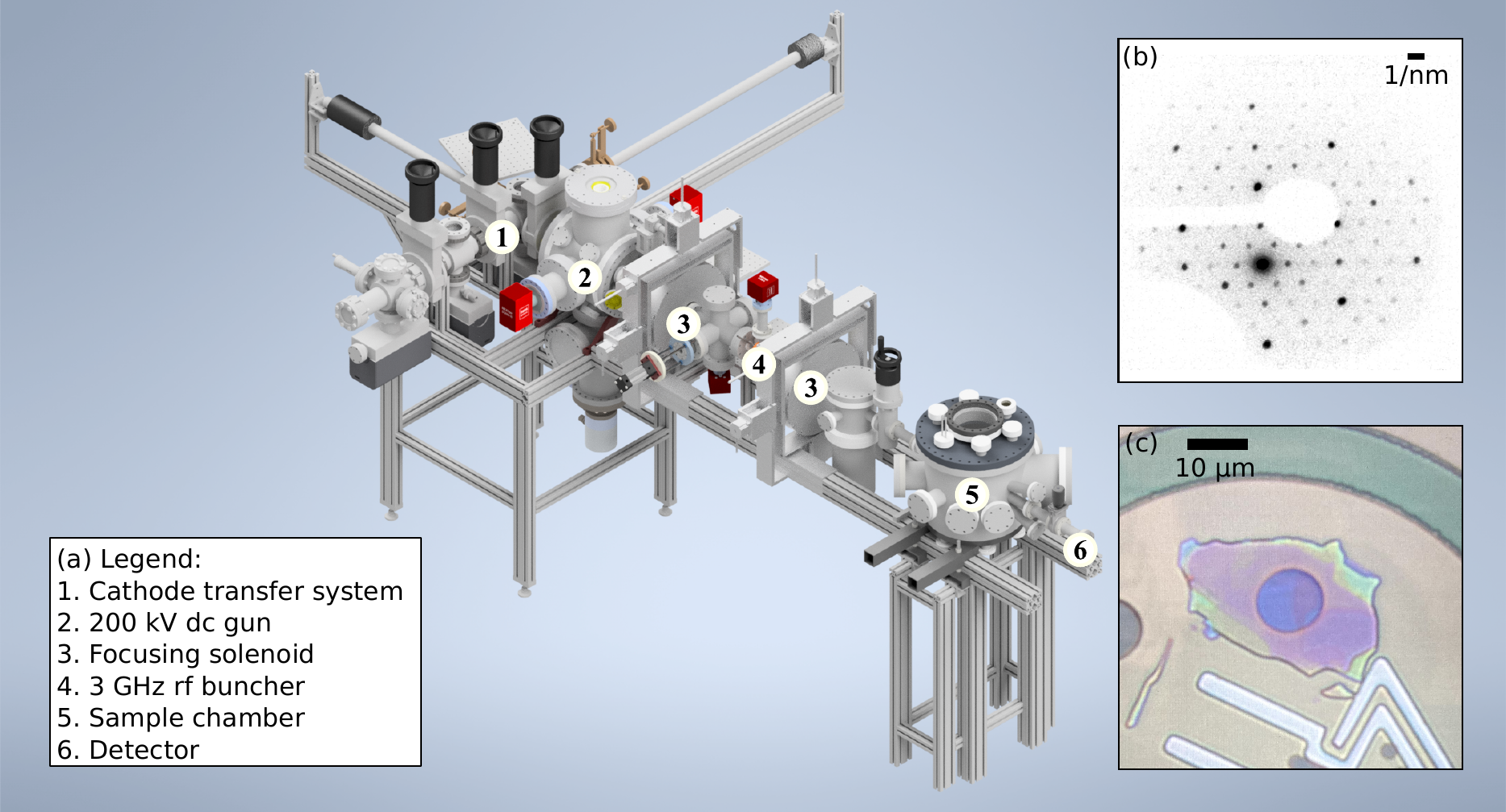}
    \caption{(a) Schematic of the UED apparatus. The beamline consists of a 200 kV dc gun, two solenoids, a 3 GHz rf bunching cavity, the sample chamber, and a detector. The total length of the beamline is approximately 2.2 m. (b) Example data taken with the apparatus, showing a selected area electron diffraction pattern produced by a single flake of $\text{Nb}_3\text{Br}_8$ \cite{bianco2020stable}.  (c) The sample flake viewed under an optical microscope, the dark oval is a 10 $\mathrm{\mu m}$ window through the SiN substrate.}
    \label{fig:beamline}
\end{figure*}

\begin{figure*}[t]
    \centering
    \includegraphics[width=\linewidth]{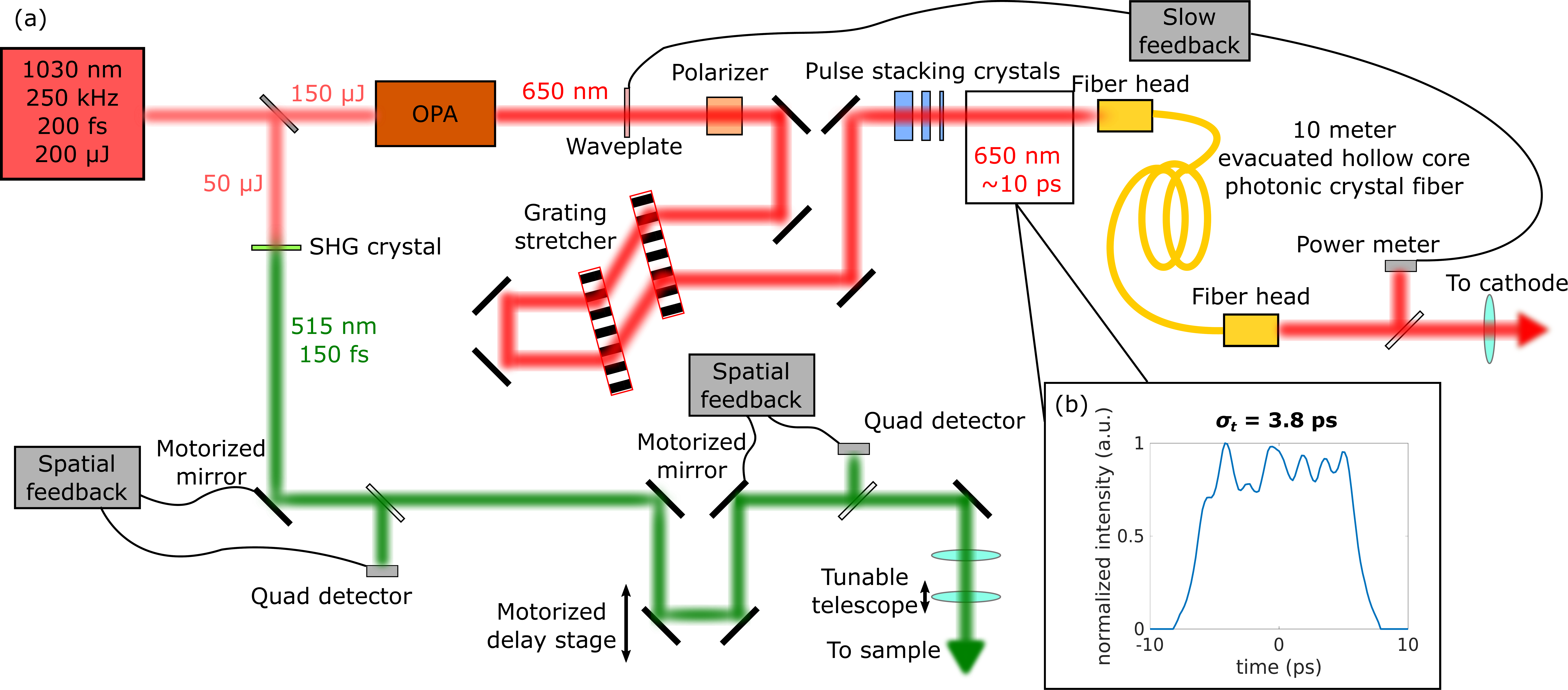}
    \caption{(a) A schematic of the laser system for the MEDUSA beamline. A 1030 nm fiber laser is used to generate both the 515 nm pump and the 650 nm photoemitting laser. The photoemitting laser is stretched with pulse stacking crystals and transported through a fiber to the cathode, to ensure position stability. The pump laser utilizes multiple feedbacked mirrors to ensure position stability. A motorized delay stage controls the relative time of arrival of the pump and probe. (b) Temporal distribution of the photoemitting laser as measured by deflecting a zero charge emitted electron beam.}
    \label{fig:laser_schematic}
\end{figure*}

\section{System Description} \label{sec:system}
Our photoemission source is a Na-K-Sb cathode grown in the Cornell photocathode lab \cite{cultrera_alkali_2016} and transported to the electron gun via a UHV suitcase. Na-K-Sb photocathodes have simultaneously high quantum efficiency ($\sim 10^{-3}$) and low MTE ($<50$ meV) when illuminated with red (650 nm) light  \cite{maxson_measurement_2015}. Our photocathode geometry is planar, and the photocathode film diameter is $\sim 1 $ cm. Thus, the profile of the photoemitting laser determines the electron source size. The photocathode is mounted on a custom INFN/DESY/LBNL-style miniplug which allows laser illumination in one of either transmission or reflection mode \cite{wells_mechanical_2016, karkare_temperature-dependent_2017}. In reflection mode, we use a laser spot size of 25 $\mathrm{\mu}$m rms, while in transmission mode, a final lens placed close to the cathode forms a minimum laser spot size of 5 $\mathrm{\mu}$m rms. The required pointing stability of the photoemitting laser is much stricter in transmission mode than in reflection mode; the results reported in this paper were obtained in reflection mode. A dc gun accelerates the beam \cite{lee_cryogenically_2018} to a maximum 200 keV; our typical operating energy is 140 keV. A schematic of the beamline is shown in Fig. \ref{fig:beamline}. Following the dc gun, the beam is focused transversely by two solenoid lenses, and longitudinally by a 3 GHz bunching cavity. We place the transverse and longitudinal foci in the plane of the sample, and the diffraction pattern is transported by a drift to a modular detector section of variable length, depending on the reciprocal space field of view required.

The laser system for generating both the pump and probe beams is a 250 kHz, 1030 nm, Yb fiber laser (Amplitude Syst\`emes Tangerine). One quarter of the available 200 $\mu$J pulse energy is taken to generate the pump (typically 515 nm, via second harmonic generation), while the rest drives an optical parametric amplifier (OPA, Amplitude Syst\`emes Mango) to generate tunable visible light for photocathode illumination. A schematic of the entire setup is shown in Fig. \ref{fig:laser_schematic}. The OPA is tuned to generate 300 fs pulses at 650 nm, with 10 $\mathrm{\mu}$J pulse energy. A grating pair stretches the 300 fs pulse to approximately 1 ps, and the pulse is further broadened to 10 ps fwhm via temporal stacking crystals \cite{bazarov_shaping_2008}. This long pulse length is chosen to obtain a ``cigar'' electron bunch aspect ratio upon photoemission, which is known to alleviate transverse space charge forces \cite{daniele_max_current_2014, renkai_nanometer_2012}. Furthermore, the optimized simulations, shown in Fig. \ref{fig:pareto} and described in more detail in section \ref{sec:spatial}, all required pulses with fwhm $>5$ ps. We choose stacking crystals over a single high-dispersion grating stretcher, as large dispersion combined with the inhomogenous spectrum of the OPA would introduce temporal profile distortions.

Multiple feedback mechanisms ensure laser power and pointing stability. The photoemitting laser intensity is set by a waveplate controlled by a slow ($\sim5$ Hz) feedback mechanism, which primarily corrects for thermal drifts. The photoemitting laser is transported to the cathode via a single mode hollow core fiber, which both ensures position stability and cleans the spatial mode of the OPA. The pump laser position and angle are read by two quadrant detectors, each of which controls a fast piezo-mirror that holds the position of the laser on the sample constant to 1 $\mu$m rms. This position stability is of critical importance given the small size of both pump and probe. The pump laser size can be tuned down to $<10$ $\mu$m  rms at the sample in the current optical configuration. 

The bunching cavity is a 3 GHz reentrant TM01 cavity based on the Eindhoven design \cite{van_oudheusden_electron_2007}. The low-level 3 GHz rf signal is derived from the 50 MHz laser oscillator; the beam dynamics are therefore largely insensitive to laser oscillator phase drift. A fast feedback system based on the work of Otto et al.\cite{otto2017solving} controls both both the phase and the amplitude of the buncher rf field as measured by a field probe on the cavity.

\begin{figure}
    \centering
    \includegraphics[width=1.0\linewidth]{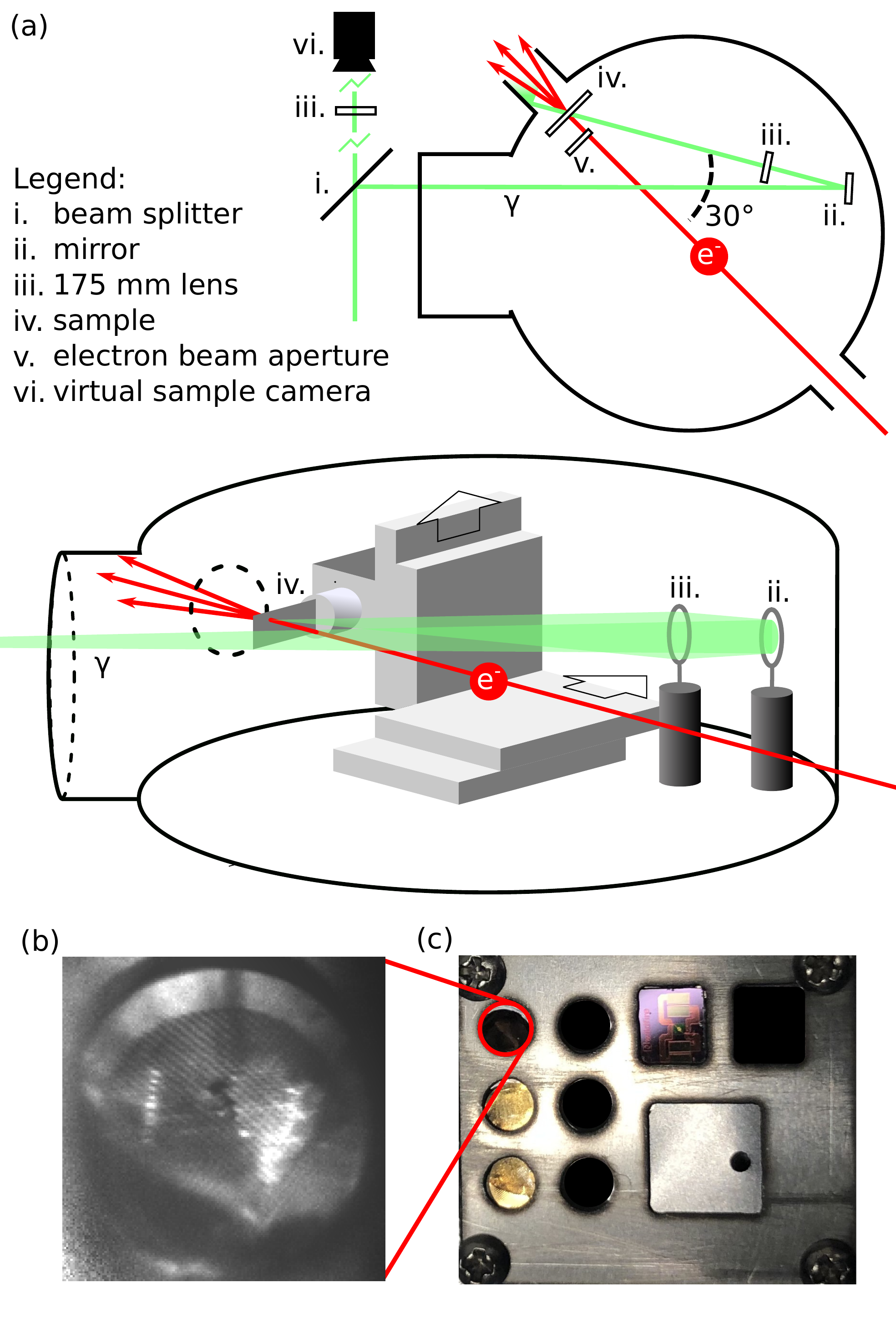}
    \caption{(a) Top down illustration of our sample chamber, configured for ultrafast electron diffraction.  We split the pump beam at the chamber entrance. The main pump path reflects off a final mirror and passes through a final focusing lens before reaching  the sample. An aperture collimates the electron probe. The pump path bypasses the electron aperture and makes a $30^\circ$ angle with the electron path. The other pump path, split at i., is focused by an equivalent lens onto a camera we use to track the pump position on the sample. (b) A view of a copper TEM mounted in the sample holder inside the chamber from a camera located outside the vacuum. (c) The sample holder accommodates 3 mm TEM grids, among other standard TEM sample mounts, as well as knife edges and pinholes for measuring emittance. The sample can be remotely translated transversely and up to two tilt axes.}
    \label{fig:chamber}
\end{figure}

A large volume sample chamber provides flexibility in experimental design. Fig.~\ref{fig:chamber}(a) is a diagram of the elements of the sample chamber when configured for ultrafast electron diffraction experiments. Pump pulses enter through a chamber window and a final, in-vacuum mirror determines the position of the pump spot on the sample.  A 175 mm lens downstream of the final mirror focuses the pump to form a waist at the sample. An out-of-vacuum, remotely adjustable lens doublet enables tuning of the final spot size on the sample. Prior to entering the chamber, a beam splitter sends part of the pump pulse energy to an out-of-vacuum camera ---the virtual sample camera--- that monitors the position of the pump pulse on the sample: the path-length to the camera is equal to that of the sample from the splitter, and a lens of equal focal length focuses the picked-off pulse to a waist at the camera CCD. Another camera looking into the sample chamber monitors the position of the sample stage, as well as the physical condition of the sample. Fig.~\ref{fig:chamber}(b) shows an example image of a 3 mm copper TEM grid (used for the destructive plasma timing diagnostic discussed below), with pump laser damage clearly visible. 

The flexibility to modify the sample setup entails routine vacuum venting of the sample chamber, making a bakeout impractical. Therefore, the vacuum in the sample chamber remains in the mid $10^{-8}$ Torr level during beam running. To prevent poisoning of the sensitive alkali antimonide photocathode, a vacuum conductance aperture and a large non-evaporable getter pump ($\sim$ 1000 l/s pumping speed) separate the beamline into two portions:  sample chamber plus detector ($10^{-8}$ Torr) and the ultra-high vacuum portion, consisting of the gun itself ($<10^{-11}$ Torr) and the beam optics section ($10^{-10}$ Torr). The resulting operational lifetime of our photocathodes is several months.

We place the probe-defining aperture 10 - 15 mm upstream of the sample. This stand-off distance allows the pump beam to reach the sample without clipping on the electron aperture while minimizing the contribution of momentum spread to the probe size on the sample. Pump and probe rays make an angle of $30^\circ$ at the sample plane. Pulse front tilt does not significantly change the pump pulse length because of the pump's small transverse size: the transverse full width at half max is one third the pulse length. 

We have tested two detector modalities. For emittance measurements, or for diffraction measurements where high momentum resolution is required, we use a high spatial resolution 50 $\mu$m thick Ce:YAG scintillator screen coupled to a cooled, scientific CMOS camera (Teledyne Photometrics Prime BSI-Express) with a lens capable of 1:1 image magnification. For most diffraction experiments, even higher collection efficiency is desired. In this case we use a P11 phosphor with a high numerical aperture lens system on the camera. The detectors are a separate vacuum module from the sample chamber, so they can be exchanged, or the camera length can be modified, without breaking the chamber's vacuum. 

\begin{figure*}
    \centering
    \includegraphics[width=1.0\linewidth]{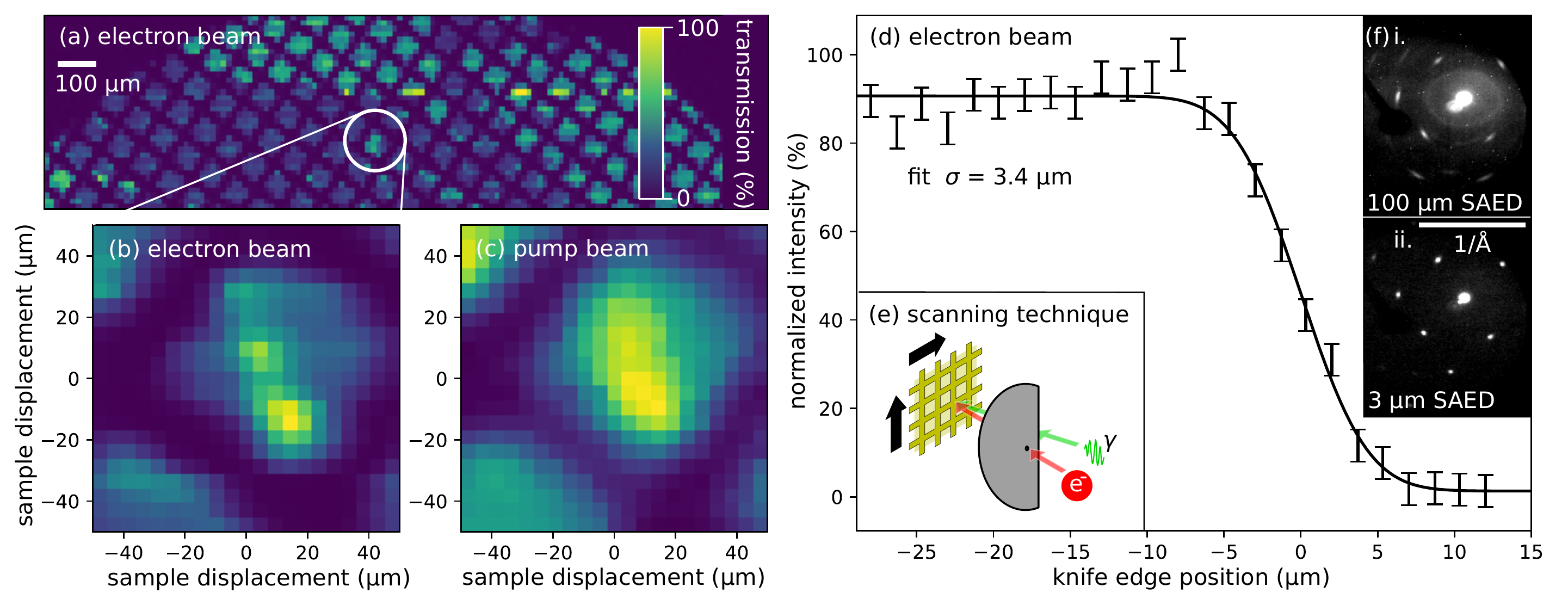}
    \caption{(a) Scanning electron transmission image of our ultrafast diffraction sample, a gold film mounted on a 3 mm gold TEM grid. (b) Finer electron beam scan of a unique feature in the sample film. (c) Transmission scan of the same feature with the pump laser beam, for the purpose of verifying alignment of electron beam and pump.  (d) Beam size measured at the sample plane through the final collimating aperture. There is a silicon knife edge in the sample plane, and the beam size is measured by fitting an error function to the intensity as a function of the position of the knife edge. The beam size is measured to be 3 microns rms. (e) Illustrating the two transmission modes, in red the electron beam is collimated by a 10 $\mu$m diameter aperture, in green the pump beam is steered around the electron aperture, making an angle with the electron beam of 30$^\circ$ at the sample plane; we perform the scan by moving the sample, shown in gold. (f) Selected area electron diffraction of the same gold sample comparing i. a 100 $\mathrm{\mu m}$ rms probe size against ii. a 3 $\mathrm{\mu m}$ rms probe size.}
    \label{fig:knife}
\end{figure*}

\begin{table}[h]
    \centering
    \begin{tabular}{c|c}
    \hline
          Beam energy & 140 keV  \\
          Cathode spot size & 25 $\mathrm{\mu m}$ rms \\
          Charge on target  & Up to 0.1 fC \\
          Probe size on target & 3.5 $\mathrm{\mu m}$ rms \\
          Normalized emittance & $\leq$1 nm-rad \\
          Bunch length & $<200$ fs rms \\
          Pump fluence & Up to 1 J/cm$^2$ \\
          Pump size on target & Down to 10 $\mathrm{\mu m}$ rms \\
    \hline
    \end{tabular}
    \caption{Routine beamline parameters for ultrafast micro-diffraction experiments.}
    \label{tab:beamline}
\end{table}

\section{Spatial Resolution}\label{sec:spatial}

Spatial resolution in this beamline can be characterized by both the reciprocal space resolution and the beam size in real space. High reciprocal space resolution results in sharp diffraction peaks, while a small beam size enables the beam to probe small spatial features in the sample.

Reciprocal space resolution $\Delta s$ can be expressed in terms of the rms spot size on the sample $\sigma_x$ and normalized emittance $\epsilon_n$ as \cite{weathersby_mega-electron-volt_2015}:
\begin{equation} \label{eq:momentum_transfer}
    \Delta s = \frac{2 \pi}{\lambda_e} \frac{\epsilon_n}{\sigma_x},
\end{equation}
where $\lambda_e$ is the electron de Broglie wavelength.

As shown in Eq. \ref{eq:momentum_transfer}, there is a tradeoff between reciprocal space resolution and beam size. The beam size is primarily determined by the size of the probe-defining aperture and the size of the target sample, so the reciprocal space resolution in this system can only be improved by lowering the transverse emittance. Thus, it is critical to measure the emittance to quantify the performance of this apparatus, in both the high-charge and micro-diffraction modes of operation.

In the high-charge mode, the probe-defining aperture is removed, and the typical charge is up to $10^5$  electrons/bunch, or 16 fC, at the sample plane. This is also the charge per bunch typically delivered to the aperture in micro-diffraction experiments. We measure the emittance of this bunch by first bringing it to a waist at the sample. Next, scanning the 10 micron diameter aperture across the beam both horizontally and vertically, and imaging the transmitted beam distribution on a viewscreen (which in this case represents the sample momentum distribution), we generate a full 4d transverse phase space of the beam. We then directly compute the emittance by calculating the sigma matrix of the beam. With this method, we measure a projected, normalized emittance of 13 nm-rad at 16 fC with an rms beam size of 43 $\mathrm{\mu m}$. Achieving this value required not only the correction of normal quadrupole and skew-quadrupole, but also sextupole components in the transverse and longitudinal focusing optics with dedicated quadrupole and sextupole corrector magnets just downstream of the second solenoid. This correction procedure will be described in a forthcoming manuscript, along with the complex space charge dynamics and compensation involved.

The micro-diffraction mode, the routine configuration for diffraction from small crystal flakes, e.g. panel (c) of Fig. \ref{fig:beamline}, uses the probe-defining aperture to generate very small probes. Typical beamline operational parameters in this mode are listed in Table \ref{tab:beamline}. The right side of Fig. \ref{fig:knife} shows a knife edge scan performed at the sample plane, yielding a transverse rms probe size of 3 microns. The bunch charge through the aperture is ~500 electrons, and the normalized emittance is $0.7  \pm 0.1$ nm-rad. The emittance in this case was measured by bringing the beam to a focus on the pinhole, measuring its size at the sample with the knife edge and measuring the momentum spread using the beam size on the final screen.

\begin{figure}
    \centering
    \includegraphics[width=0.5\textwidth]{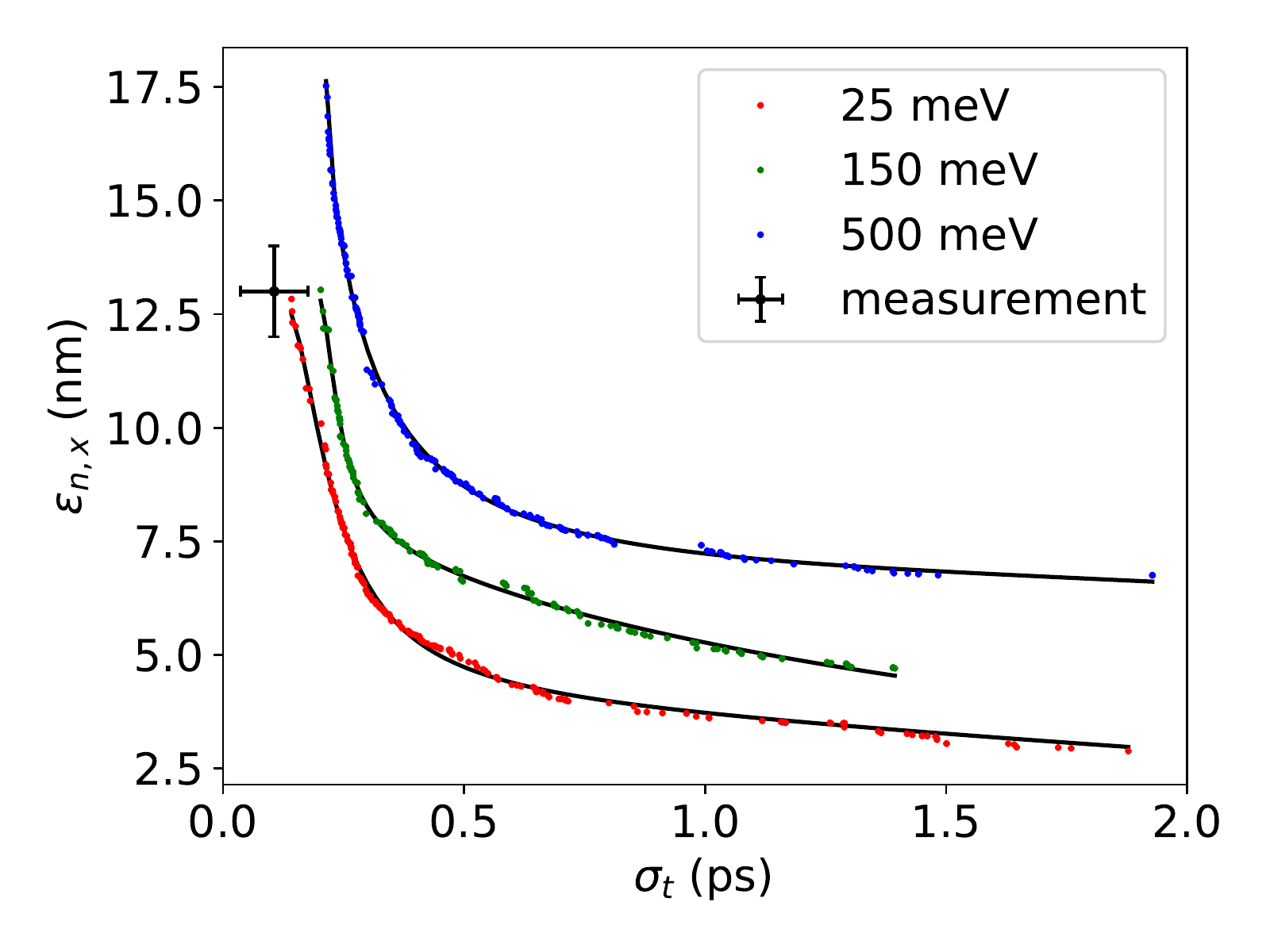}
    \caption{Pareto fronts generated in simulation by a genetic algorithm optimizer for cathode MTEs of 25 meV (red), 150 meV (green), and 500 meV (blue) with a final bunch charge of 16 fC. The optimizer was set to minimize emittance and bunch length, holding MTE, gun voltage, and element positions constant while being free to pick operational parameters. The measured emittance and bunch length are consistent with an MTE less than 150 meV. The solid lines are included solely for the purpose of guiding the eye.}
    \label{fig:pareto}
\end{figure}

In order to understand the role of low MTE in our space charge dominated conditions, we use General Particle Tracer (GPT) \cite{gpt}, a particle tracking PIC code, to simulate the changes in beam dynamics for various MTEs. GPT is coupled with a multiobjective genetic algorithm \cite{gulliford_multiobjective_2016}, which is set to optimize both bunch length and emittance at the sample plane, generating a curve of optimal emittance for each bunch length, known as a Pareto front. We fix the gun voltage and element positions and allow the electron optics and 3d laser pulse shape to vary, as would be the case in operation. Pareto fronts generated for MTEs of 25, 150, and 500 meV are shown in Fig.~\ref{fig:pareto}, and in these optimizations, the optimizer was permitted to emit up to 160 fC at the cathode and is required to transmit 16 fC to the sample after traversing physical apertures (for example, the bore of the buncher and pipe) in the beamline. We see a reduction in emittance of roughly 30\% from 150 meV to 25 meV and roughly a factor of 2 from 500 meV to 25 meV, averaged over all displayed bunch lengths.

From a physical standpoint, the relationship between MTE, brightness, and bunch length can be understood by considering the emitted charge. To achieve the same final emittance and charge at higher MTE, simulations show that it is necessary to emit a higher charge density and select out a smaller fraction of the total beam. Higher charge density results in larger longitudinal space charge forces prior to the sample plane, leading to longer bunch lengths.

Comparing the measured emittances to simulation, the optimal simulated emittance for a 200 fs bunch length is 17.5 nm-rad at an MTE of 500 meV, 13 nm-rad at 150 meV, and 10 nm-rad at an MTE of 25 meV. As will be covered in section \ref{sec:temporal}, the bunch lengths we measure are lower than 200 fs, so the measured high-charge emittances are consistent with an MTE substantially less than 150 meV.

To help locate our machine in the parameter space of other ultrafast electron instruments, we compute 4d and 5d beam brightnesses from our beam diagnostic measurements. It is convenient to define the 5d normalized beam brightness $B_{np}$ in terms of normalized emittance and peak current $I_p$ as
\begin{equation} \label{eq:brightness}
 B_{np} := \frac{I_p}{\pi^2 \epsilon_n^2}.
\end{equation}
We define the peak current of a beam with charge per bunch $Q$ and rms bunch length $\sigma_t$ as
\begin{equation} \label{eq:current}
    I_p := \frac{Q}{\sqrt{2\pi} \sigma_t}.
\end{equation}
In the high-charge mode, the 5d brightness at the sample plane is $4\times10^{13} \mathrm{A/m^2\text{-}rad^2}$, and $7\times10^{13} \mathrm{A/m^2\text{-}rad^2}$ in the micro-diffraction mode, consistent with the expectation that core brightness is higher than average brightness. For machines operating in the same charge and emittance regime, the natural units of 4d brightness, $Q/\epsilon_n^2$, are electrons per nanometer-radian squared per pulse. Our machine delivers 600 electrons/$(\text{nm-rad})^2$ in high-charge mode and 1000 electrons/$(\text{nm-rad})^2$ in micro-diffraction mode at the sample plane. For experimental applications, brightness at the sample plane is the relevant figure, not source brightness, as space charge forces and lens aberrations cause brightness loss in beam transport \cite{ piazza2013design, bach2019coulomb, curtis2021toward}.

It is informative to compare the brightness of MEDUSA with MeV-UED beamlines and ultrafast electron microscopes (UEMs) equipped with nano-tip sources, as this subset of ultrafast electron technology spans multiple orders of magnitude in charge per bunch, repetition rate, emittance, and temporal resolution around our operating point. Additionally, the brightness values of these machines have been well-characterized. Megavolt diffraction beamlines have demonstrated 5d brightnesses in the range of $10^{12}$ to $10^{14}$ $\mathrm{A/m^2\text{-}rad^2}$ and 4d brightnesses in the range of 10 to $10^3$ electrons/$(\text{nm-rad})^2$ in both high-charge and micro-diffraction modes \cite{weathersby_mega-electron-volt_2015, shen2018femtosecond, zhu2015femtosecond, filippetto2016design}. On the other end of the charge spectrum, at less than one electron per pulse delivered to the sample, nanotip UEMs achieve 5d brightnesses in the range of $10^{13}$ to $10^{15}$ $\mathrm{A/m^2\text{-}rad^2}$ and 4d brightnesses in the range of $10^3$ to $10^4$ electrons/$(\text{nm-rad})^2$ when operated in high coherence mode \cite{houdellier2018development, ropers_utem_2017}. Nanotips have been shown to yield $10^{12}$ $\mathrm{A/m^2\text{-}rad^2}$ and $10^2$  electrons/$(\text{nm-rad})^2$  at charges greater than one electron per pulse delivered to the sample; however, owing to brightness loss, users do not typically operate nanotip UEMs at these charges  \cite{ropers_utem_2017}.

This beamline occupies an unfilled position in parameter space, and is complementary to other state-of-the-art time-resolved electron scattering devices. Looking to the examples cited above, this beamline achieves comparable transverse brightness to existing MeV diffraction beamlines, but with high repetition rate and reciprocal space resolution in a compact footprint. However, compared to the same MeV devices, this beamline has coarser temporal resolution and more stringent vacuum requirements, and more complicated space charge compensation. Compared to nanotip UEMs, this beamline offers finer temporal resolution and much larger charge per pulse at the cost of lower transverse brightness.

As stated in the introduction, real space resolution, i.e., probe size, is a critical figure of merit for this beamline. In addition to the aforementioned knife edge scans, the left side of Fig. 4 provides another demonstration of the real space resolution of the electron beam. By scanning the position of the sample, we make a transmission image of the sample. Our target is a gold TEM grid with bar width of 25 microns and a pitch of 83 microns. We resolve fine details of this target using the electron beam scan with better resolution than the equivalent scan with the pump laser, which was measured to have a size of 10 microns rms. Not only does this method provide micron-resolution imaging, it also allows us to align pump and probe with single micron precision.

\section{Temporal Resolution}\label{sec:temporal}
Two terms add in quadrature to determine the temporal resolution of the instrument: (i) the stability of the gun and bunching fields, and (ii) the bunch length of the probe beam. A compact 3 GHz TM110 deflecting cavity (DrX Works) placed very near the plane of the sample transversely streaks the electron beam and provides direct measurements of both (i) and (ii).

\begin{figure}
\centering
    \includegraphics[width=\linewidth]{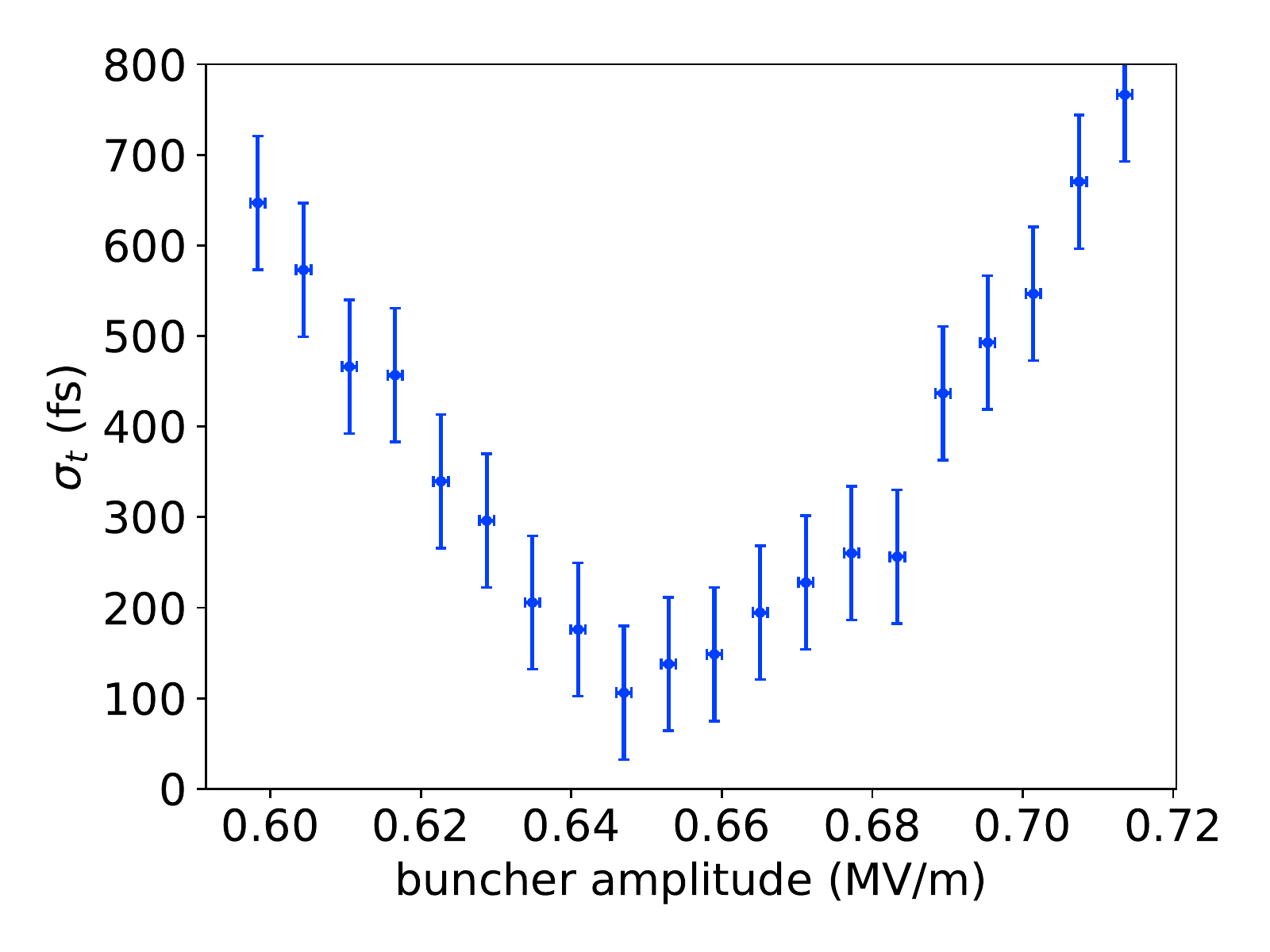}
    \caption{RMS bunch length as a function of buncher field amplitude. The minimum bunch length measured is less than 200 fs. The displayed error bars represent the mean of the individual standard errors for each measurement.}
    \label{fig:bunch_length}
\end{figure}

\begin{figure}[h]
    \centering
    \includegraphics[width=1.0\linewidth]{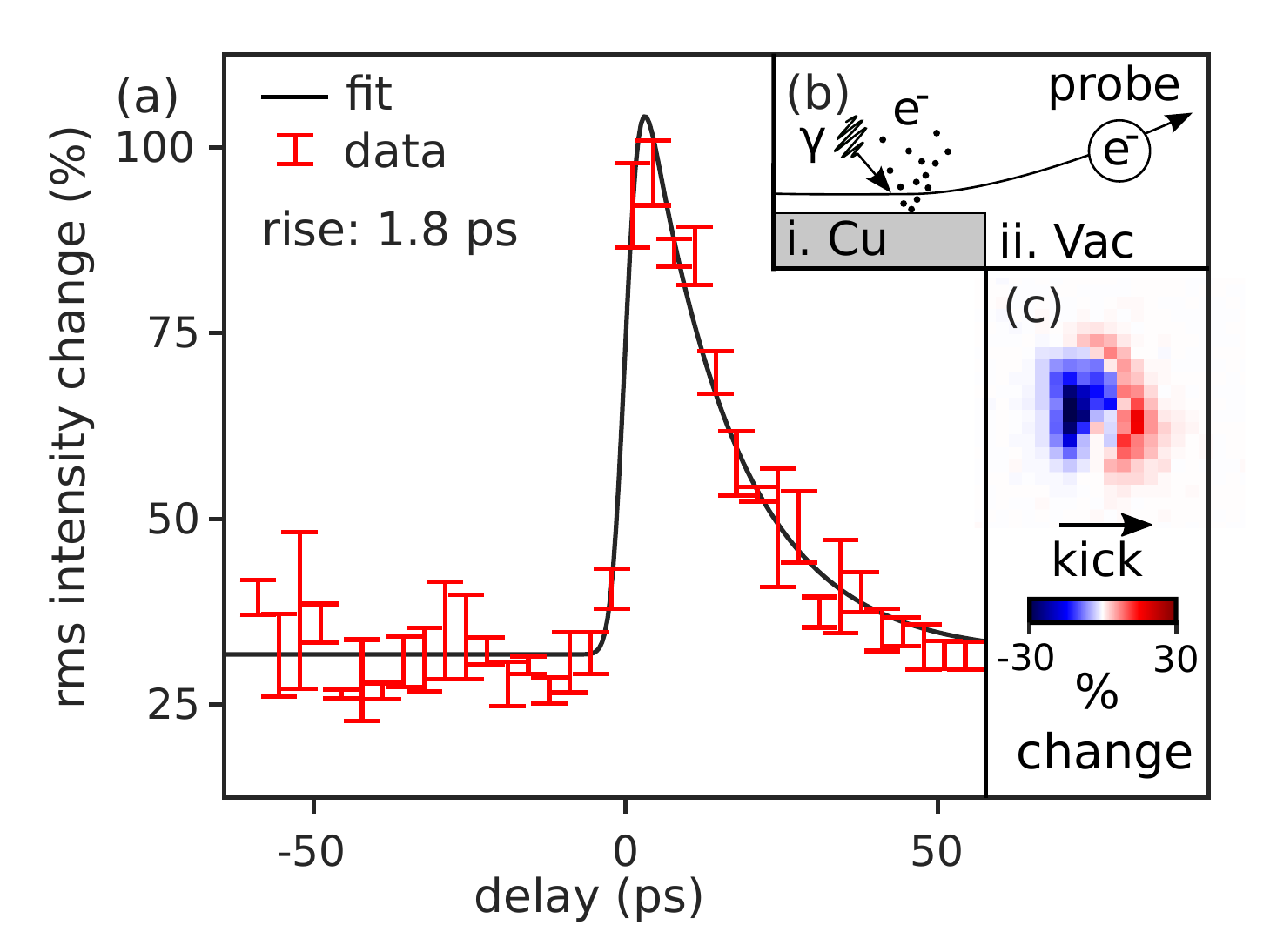}
    \caption{(a) Measurement of time-dependent photoemitted plasma lensing, showing the rms per pixel change in beam intensity as a function of time. (b) Schematic of the mechanism that drives the beam response, i. pump laser photoemits an electron gas from a copper grid, ii. Coulomb field from emitted gas deflects the electron beam. (c) Example of the detected difference image used to compute the signal plotted in (a).}
    \label{fig:plasma}
\end{figure}

\begin{figure*}
    \centering
    \includegraphics[width=1.0\linewidth]{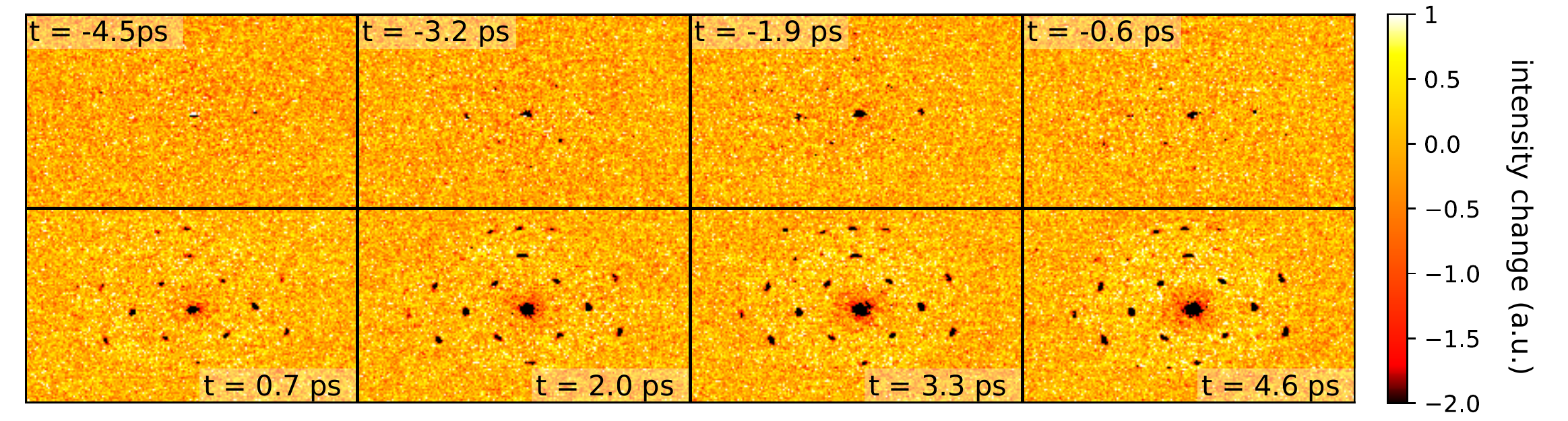}
    \caption{Time-dependent diffraction from a gold sample pumped by 515 nm light. Each panel is a composite of 100 acquisitions at a 5 s exposure time and 5 kHz electron beam repetition rate, taken at a fixed pump delay indicated by the inset text.}
    \label{fig:frames}
\end{figure*}

The change in beam centroid as the deflector kicks the beam gives the bunch time of arrival. After subtracting off the transverse jitter in quadrature, we measure the time of arrival jitter to be 170 fs. We believe this to be an upper bound, as it includes contributions from deflector phase jitter. We estimate bunch length by deconvolving the beam width with the deflector on from the beam width with the deflector off. By fitting for the kernel of the convolution, we extract the contribution to the beam width from the deflected bunch length. Fig. \ref{fig:bunch_length} shows measured bunch length as a function of buncher field amplitude. The minimum bunch length is measured to be $110 \pm 70$ fs rms, with the majority of the uncertainty arising from the deconvolution of the transverse beam size from the deflected bunch length, as well as deflector phase jitter. We plan to improve our measurements of time of arrival jitter by performing streaked diffraction experiments \cite{liNoteSingleshotContinuously2010, musumeciCapturingUltrafastStructural2010} which will provide a time-stamped signal to reconstruct beam time of arrival, mitigating the effects of deflector phase jitter. To generate an estimate of the overall temporal resolution, we add the measured time of arrival jitter in quadrature with the bunch length to yield $200 \pm 40 $ fs rms.  

A variable path length controls the time of arrival of the pump pulse at the sample plane. Synchronizing the pump and probe beams requires a target sample that responds promptly to the pump pulse. A standard technique is to aim the pump at a copper TEM grid and photoemit an electron gas \cite{scoby_effect_2013}, recording a time-resolved shadowgram of the electron gas with the probe beam. At a pump wavelength of 515 nm, the emitted charge is quadratic in pump energy and the effect on the probe becomes obvious at fluences above $100 \ \mathrm{mJ/cm}^2$. Fig.~\ref{fig:plasma}(a) shows the signal we measure upon scanning pump delay. Fig.~\ref{fig:plasma}(b) sketches the physical mechanism: the ejected plasma's coulomb field deflects the probe beam. The size of the deflection (exaggerated in Fig.~\ref{fig:plasma}(b)), though small compared to the beam size, is readily apparent in a difference image comparing the probe beam on the final screen between pump on and off. Fig.~\ref{fig:plasma}(c) shows an example difference image. The signal we compute is the rms intensity of these difference images, cropped to a region of interest, as a function of delay time.

\section{Ultrafast Electron Diffraction}\label{sec:ued}

As a proof-of-principle experiment, we measure the ultrafast heating of a gold diffraction sample (TED Pella \#646) with mosaicity observed in static diffraction patterns indicating domains of sizes $\sim$ 10 - 100 $\mathrm{\mu}$m. According to a two-temperature model of this process, the 515 nm pump beam deposits its energy into conduction electrons, which as a subsystem thermalize on a time scale much shorter than the 200 fs pump pulse length. The hot electron gas then equilibrates with the lattice over a few picosecond relaxation time $\tau$. The transverse profile of the deposited energy is Gaussian with an rms size of $12 \ \mathrm{\mu m}$. This pump size represents a trade-off between uniform illumination and minimizing the total deposited energy. The probe optics are chosen to maximize transmission through the collimating aperture, producing an rms probe size of $5 \ \mathrm{\mu m}$. Hence, the sample temperature in the probed region is close to spatially uniform. Measuring the incident, transmitted, and reflected pump power {\em in situ}, we estimate the absorbed pump fluence to be no greater than $0.8 \ \mathrm{mJ/cm}^2.$ 

In the data-set shown in Fig.~\ref{fig:frames}, we scan delays at $1.3$ ps intervals. Each panel in the figure shows the difference in intensity between pump on and pump off on the detector screen, averaging $1.25\times 10^6$ pulses per delay stage position. We alternate acquisitions of $5 \ \mathrm{s}$ at a repetition rate of $5 \ \mathrm{kHz}$ between pump-on and pump-off to control for slow drifts in the primary beam current.

\begin{figure}[b!]
    \centering
    \includegraphics[width=1.0\linewidth]{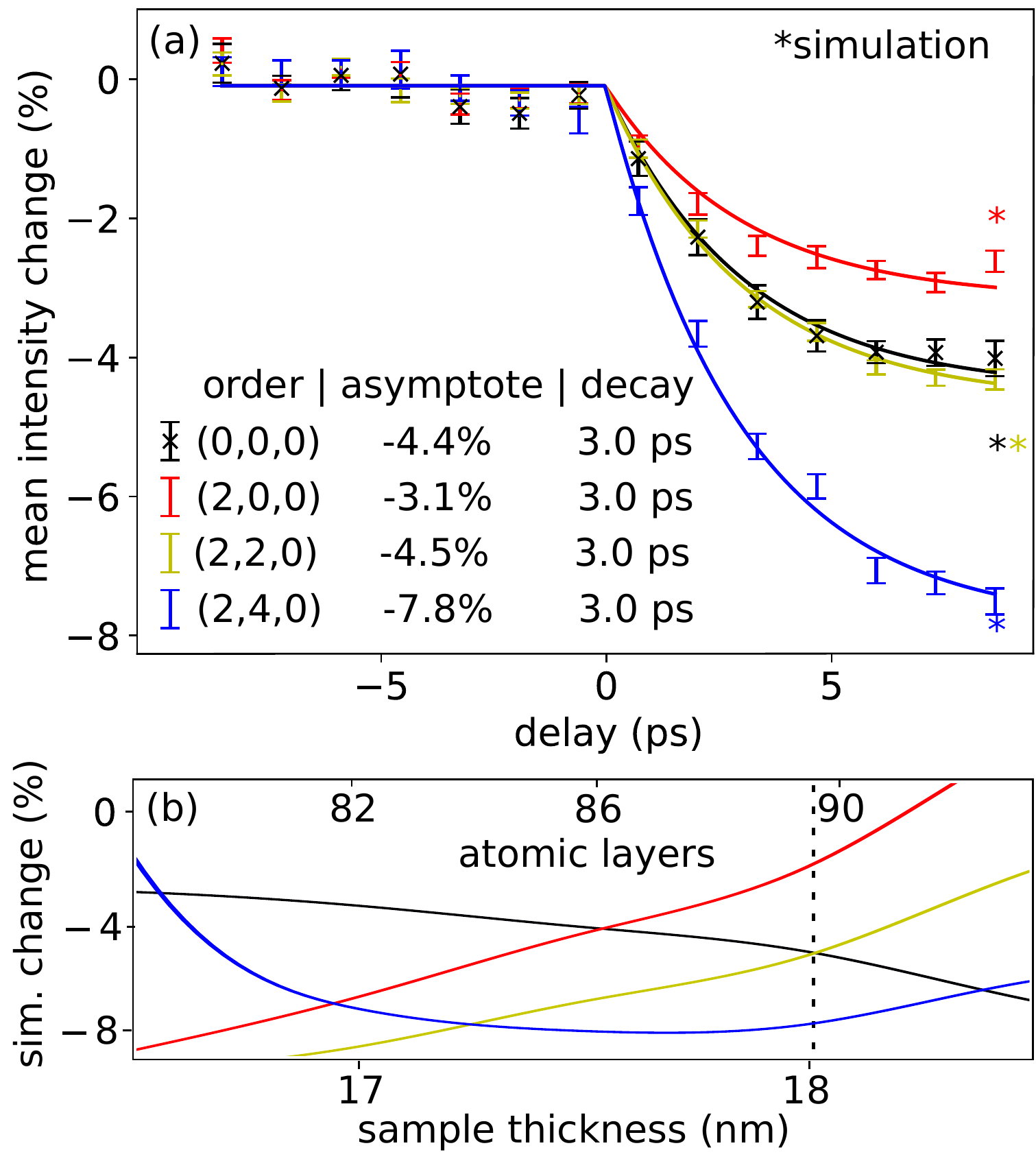}
    \caption{Bragg peak intensity as a function of pump delay. Peak intensities are estimated from the data shown in Fig.~\ref{fig:frames} by taking the maxima of Gaussian filtered images. The quantity plotted is the change in this intensity normalized by its fully-relaxed (pump-probe delay 100 microseconds) value. (b) Simulated signal as a function of sample thickness including multiple scattering effects, assuming a cold temperature of 273 K and a hot temperature of 423 K; colors have the same meaning as in panel (a), dashed line indicates a thickness that reproduces qualitatively the dependence on scattering angle in the experimental data, values along this line are plotted as asterisks in panel (a).}
    \label{fig:DW}
\end{figure}

Grouping peaks related by lattice symmetries and averaging the change in peak intensity at each delay time gives the results shown in Fig. \ref{fig:DW}(a). To each data set we fit an exponential decay, constraining the fit to find a common decay time $\tau$ for all curves and obtaining $\tau = 3.0 \pm 0.3 \ \mathrm{ps}$. 

\begin{figure}[t!]
    \centering
    \includegraphics[width=1.0\linewidth]{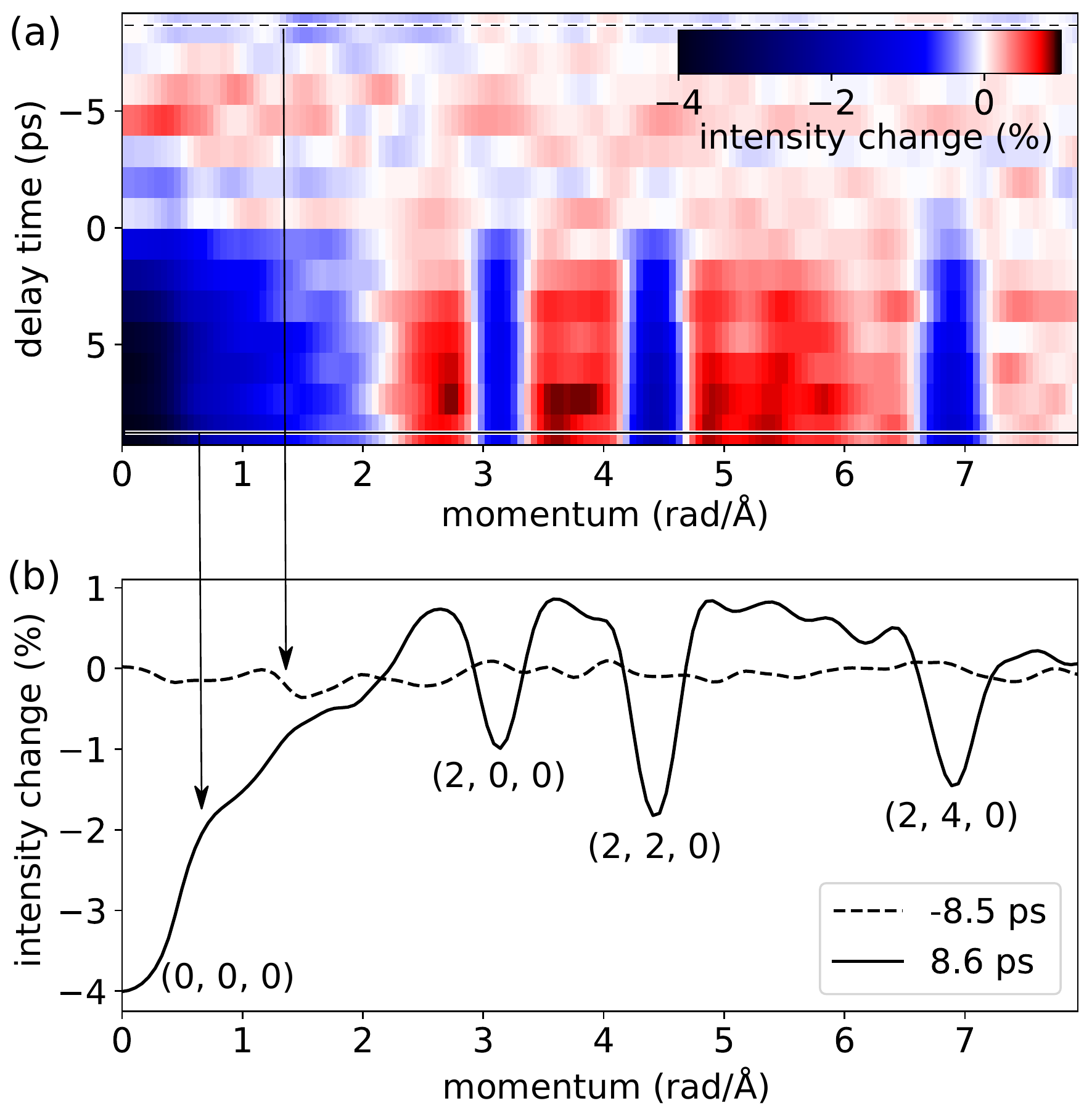}
    \caption{(a) Scattering intensity as a function of the magnitude of the scattering vector and pump delay time; the data are calculated by averaging the images in Fig.~\ref{fig:frames} over the azimuthal coordinate, centered on the primary peak. (b) A comparison of two cross sections of panel (a), showing explicitly the percentage change in scattering intensity as a function of the magnitude of the scattering vector; the dashed curve corresponds to the cross section indicated by a dashed line in panel (a), likewise the solid curve.}
    \label{fig:radial}
\end{figure}

Averaging the diffraction data along the azimuth provides a complementary 2D visualization of the ultrafast time-dependent effect. Fig.~\ref{fig:radial}(a) shows delay time on the vertical axis and the magnitude of the scattering vector on the horizontal. The four dark streaks that begin to appear at $t=0$ correspond to the primary beam and the three reflections plotted in Fig.~\ref{fig:DW}. The dashed and solid horizontal lines correspond to the cross-sections plotted in Fig.~\ref{fig:radial}(b). In addition to the suppression of the diffraction peaks, Figs.~\ref{fig:radial}(a) and (b) reveal an enhancement of the diffuse scattering signal at scattering vectors in the interval between $2$ and $6 \ \text{rad/\AA}$. With an upgrade to our detector we plan to better resolve the momentum dependence of time-dependent diffuse scattering, revealing the non-equilibrium dynamics of lattice phonons \cite{chase2016ultrafast, stern2018mapping}.

The leading effect of ultrafast heating on the probe beam in the weak scattering limit is to wash out the coherence of the partial waves scattered from individual atoms. If the sample's initial temperature is well above its Debye temperature, and the temperature change $\Delta T$ is small compared to the sample's melting point,  
then the fractional suppression of the beamlet scattered by reciprocal lattice vector ${\bf k}$ is proportional to $k^2 \Delta T$. Though the gold film is approximately 20 nm thick, at our primary beam energy, multiple scattering is significant and hence the trend in Fig.~\ref{fig:DW}(a) does not show a quadratic dependence on scattering angle. 

To quantify the importance of multiple scattering, we perform electron diffraction simulations with the slice-method code, $\mu$STEM \cite{allen2015modelling}. Simulation results as a function of thickness and temperature are shown in Fig.~\ref{fig:DW}(b). The vertical axis represents the relative change in diffraction peak intensity between the two temperatures 273 and 423 K, and the horizontal axis is the simulated crystal thickness. The vertical dashed line in Fig.~\ref{fig:DW}(b) indicates a simulation thickness that reproduces the ordering of Bragg peaks seen in Fig.~\ref{fig:DW}(a). The range of the horizontal axis in Fig.~\ref{fig:DW}(b) covers 12 atomic layers. The figure thus shows that the addition of one atomic layer modulates the temperature response by more than the experimental uncertainty of the data shown in Fig.~\ref{fig:DW}(a). The simulations  agree qualitatively with our UED data, disagreeing on the scale of 10\% of the measured effect. Non-uniform specimen thickness and temperature within the probed region could plausibly account for this discrepancy. 

\section{Summary and Outlook}\label{sec:summary}
We have designed and commissioned an ultrafast electron micro-diffraction apparatus with sub-picosecond temporal resolution. Using an alkali antimonide photocathode, we generate beams with low intrinsic emittance allowing for micron scale probe size without significant loss in reciprocal space resolution. A 3 GHz rf bunching cavity compresses the beam longitudinally to a few hundred femtoseconds, while solenoids and a collimating aperture reduce the beam size to 3 $\mathrm{\mu m}$ rms. With this small probe, as a proof-of-principle experiment we showed that the apparatus can clearly resolve the ultrafast evolution of the lattice temperature of a few-micron selected area of a textured gold film.

The capability to perform ultrafast electron diffraction with single-micron scale selected areas enables the study of samples that are challenging (or impossible) to produce in larger sizes. Further, small electron probe sizes allow the commensurate reduction of pump area, which, for a constant excitation fluence, can dramatically reduce the pump energy deposition in the sample,  mitigating average heating effects and sample damage at high repetition rates.

Future planned upgrades include the installation of a direct electron detector \cite{empad}, with an expected substantial improvement in both sensitivity and data acquisition rate. Further, low MTE photoemission conditions provide electron beams with low total energy spread (potentially in the 10s of meV). With the addition of a high-resolution spectrometer, this apparatus will have the capability to probe electronic degrees of freedom on ultrafast timescales with very high energy resolution. A single probe that can correlate structural and electronic information promises to provide new insight into the role of electronic-phonon interactions in determining material properties.
\begin{acknowledgements}
 We thank Y. T. Shao and D. A. Muller for suggesting a multiple scattering interpretation of the UED data, and for introducing the authors to the $\mu$STEM software package. This work was supported by the U.S Department of Energy, grant DE-SC0020144 and U.S. National Science Foundation Grant PHY-1549132, the Center for Bright Beams.
\end{acknowledgements}
\section*{Author Declarations}
\subsection*{Conflict of Interest}
The authors have no conflicts to disclose.
\section*{Data Availability}
The data that support the findings of this study are available from the corresponding author upon reasonable request. 

\bibliography{uedperformance}

\end{document}